# Importance of Preprocessing in Histopathology Image Classification Using Deep Convolutional Neural Network

Nilgün ŞENGÖZ [1,*], Tuncay YİĞİT [2], Özlem ÖZMEN [3], Ali Hakan ISIK [4]

[1] Burdur Mehmet Akif Ersoy University, Burdur, Turkey;
0000-0001-5651-8173
[2] Süleyman Demirel University, Dept. of Computer Engineering, Isparta, Turkey;
0000-0001-7397-7224
[3] Burdur Mehmet Akif Ersoy University, Faculty of Veterinary Medicine, Dept. of Pathology, Burdur, Turkey;
0000-0002-1835-1082
[4] Burdur Mehmet Akif Ersoy University, Faculty of Engineering and Architecture, Dept. of Computer Engineering,
Burdur, Turkey; 0000-0003-3561-9375

**Abstract**

The aim of this study is to propose an alternative and hybrid solution method for diagnosing the disease from histopathology images taken from animals with paratuberculosis and intact intestine. In detail, the hybrid method is based on using both image processing and deep learning for better results. Reliable disease detection from histo-pathology images is known as an open problem in medical image processing and alternative solutions need to be developed. In this context, 520 histopathology images were collected in a joint study with Burdur Mehmet Akif Ersoy University, Faculty of Veterinary Medicine, and Department of Pathology. Manually detecting and interpreting these images requires expertise and a lot of processing time. For this reason, veterinarians, especially newly recruited physicians, have a great need for imaging and computer vision systems in the development of detection and treatment methods for this disease. The proposed solution method in this study is to use the CLAHE method and image processing together. After this preprocessing, the diagnosis is made by classifying a convolutional neural network sup-ported by the VGG-16 architecture. This method uses completely original dataset images. Two types of systems were applied for the evaluation parameters. While the F1 Score was 93% in the method classified without data preprocessing, it was 98% in the method that was preprocessed with the CLAHE method.

## 1. Introduction

In the age of Artificial Intelligence (AI), the analysis of visual information has be-come an important reality, enabling machines to process the information contained in images. In this context, it has a rapidly expanding field in many different applications such as security, transportation and health (diagnosis) in order to accelerate decision-making and analysis processes. In particular, the need for greater efficiency in image processing and analysis has become important due to the increasing volume of image data and increasingly difficult computational tasks.

Deep learning is a part of machine learning and is a special type of artificial neural network (ANN) that resembles a multi-layered human cognition system. Deep learning is of great interest today due to its use with large healthcare data. Although ANN was introduced in 1950, its training posed serious limitations and problems for solving problems due to problems such as lack of computational power and lack of sufficient data. However, given the current availability of big data today, advanced computing power with graphics processing units (GPU), and new algorithms for training a deep neural network (DNN), many limitations and problems have been resolved. In this context, deep learning approaches show impressive performances in imitating humans in various fields, including medical imaging.

Image and video processing finds wide application in medical science. Biomedical imaging techniques have an important role in both diagnosis and treatment. These techniques have significantly helped improve patients' health care. Image-guided therapy has significantly reduced the risk of human error with improved accuracy in disease detection and surgical procedures. The history of medical imaging began in the 1890s and has gradually risen thanks to today's technological infrastructure and accessible datasets.

Histopathology diagnoses diseases occurring in tissues/cells by examining tissues and cells under a microscope. In addition to diagnosing disease, histopathologists assist clinicians with patient care and management.

Digital pathology (DP) is the process by which histology slides are digitized to pro-duce high resolution images. It is becoming very popular in the field of medical imaging research. By transferring microscopic digital images to a computer pro-gram, early diagnosis and treatment methods can be developed, thus providing great benefits to both pathologists and patients. For this reason, histopathology imaging applications attract significant attention from the research community due to their rapid and effective nature. By transferring microscopic images to digital media in the laboratory environment by pathologists, the use of Artificial Intelligence algorithms in disease detection, segmentation and classification problems has been enabled.

---

*Corresponding author
*E-mail address:* nilgunsengoz@mehmetakif.edu.tr

In this study, detailed information about Paratuberculosis disease in animals will be given, and then the contribution of the CLAHE method, which is one of the preprocessing techniques used, to the learning algorithm will be examined.

## 2. Definition and Significance of the Disease

Paratuberculosis, Mycobacterium avium subsp. It is a chronic and contagious disease caused by paratuberculosis (pTB). The disease is present worldwide and mainly affects ruminant (cows, cattle, etc.) animals, but has also been found in various animal species including horses, pigs and rabbits. [2] The worst part of pTB disease is that this type, which was first encountered in 1892, unfortunately still has not been found to have a vaccine and there is no effective treatment method currently available.

The age of the animal is a very important factor in pTB. Cattle develop resistance with age and are often infected as calves. However, age-related resistance can be overcome by infectious pressure. Moreover, the rate of detectable infections in-creases with the age of the animal. These disease traits, combined with the progressive nature of pTB, mean that the age distribution of herds plays an important role in pTB dynamics. PTB causes significant economic losses. Most important ones; Substitution costs due to increased culling, increased mortality, de-creased milk production and reproductive performance, decreased slaughter value and higher susceptibility to other diseases [3].

The economic loss associated with pTB has been estimated at $20-50 per cow in infected herds. In the United States, the infection causes an annual economic loss of more than 1.5 billion dollars. In France, the average loss associated with a clinical case was EUR 1940, while each subclinical case resulted in an estimated loss of EUR 461 [4, 5]. According to a study conducted in Hungary, the economic loss from the increased culling and mortality rate of pTB seropositive cows is EUR 166 per cow. In addition to the resulting economic losses, it has been reported that the control of pTB is not only in animal health but also in the etiology of Crohn's disease seen in humans, and that pTB can be transmitted from infected animals to humans by consuming dairy products such as cheese, yogurt, cream, butter, ice cream as well as meat products [6]. However, Crohn's patients are 7.01 times more likely to detect MAP DNA compared to controls [7]. Therefore, measures to prevent MAP from entering the food chain are beneficial.

pTB is an OIE (World Organization for Animal Health) listed disease, and in a recent review of pTB control programs in 48 countries, 73% of the countries studied considered paratuberculosis to be a notifiable disease in dairy cattle. In addition, 46% of the countries surveyed in this report have an established control program. The most common reasons for having a control program include the impact of pTB on animal health, production losses, trade restrictions and animal welfare and public health concerns. The most common pTB control programs rely on a combined testing and evaluation scheme, with breaking the transmission routes within and between flocks. However, basic data on the true prevalence of pTB infection (community incidence of the disease) are needed to evaluate the effectiveness of a paratuberculosis control program [8]. Unfortunately, there is no active program against the disease in our country, and there is no union/association, etc., organized specifically for this disease, as in other countries.

With the ELISA test, the seroprevalence of pTB in the Burdur region was 6.2%, while the farm prevalence of pTB was evaluated in the study conducted in and around Burdur province, and it was determined as 58% (14/24). The prevalence of pTB in Burdur region was determined as 6.2% in Holstein cattle [9].

The aim of this study is to propose an alternative and hybrid solution method for disease diagnosis from histopathology images taken from animals with paratuberculosis and intact intestine. In detail, the hybrid method is based on using both image processing and deep learning for better results. Reliable disease detection from histopathology images is known as an open problem in medical image processing and alternative solutions need to be developed.

## 3. Material and Method
### 3.1. Overview of Deep Neural Network

Deep learning uses artificial neural networks to perform complex calculations on large amounts of data. It is a kind of machine learning that works according to the structure and function of the human brain. Industries such as healthcare, e-commerce, entertainment, and advertising often use deep learning algorithms.

While deep learning algorithms have self-learning representations, they depend on Artificial Neural Networks that reflect the way the brain computes information. During the training process, algorithms use unknown elements in the input distribution to extract features, group objects, and discover useful data patterns.

Like training machines for self-learning, this happens on multiple levels, using algorithms to build the models. Deep learning models make use of several algorithms. While no single network is considered perfect, some algorithms are better suited for performing certain tasks.

### 3.2. Convolutional Neural Network

Convolutional neural networks (CNN) have gained special status in the last few years as a particularly promising form of deep learning. Rooted in image processing, convolutional layers have penetrated almost all subfields of deep learning and are often very successful.

CNN has been applied in various applications recently due to its feature extraction, pattern detection and classification capabilities.

CNN's main architecture consists of two basic parts; a feature extractor and a classifier. The feature extractor, in turn, consists of several linked layers. CNNs consist of convolution layer (Conv), pool layers, activation function and fully connected layer [10].

Due to their ability to learn highly complex features of image data [11], CNNs have been shown in recent years to be very successful in various image classification [12-14] and object detection [15] tasks. In the current literature, the most common deep learning techniques applied to computer vision applications are based on CNNs [16].

Fully convolutional networks (FCA) are a variation of CNNs. FCA architectures only consist of convolution and pooling layers (and possible deconvolution and upsampling layers), but ultimately do not include fully connected layers. The out-put of an FCA is usually the same size as its input. For segmentation/segmentation tasks, these intensive prediction models have provided significant improvements in both efficiency and accuracy. Figure 1 and Figure 2 show example architectures for a CNN and an FCA [17].

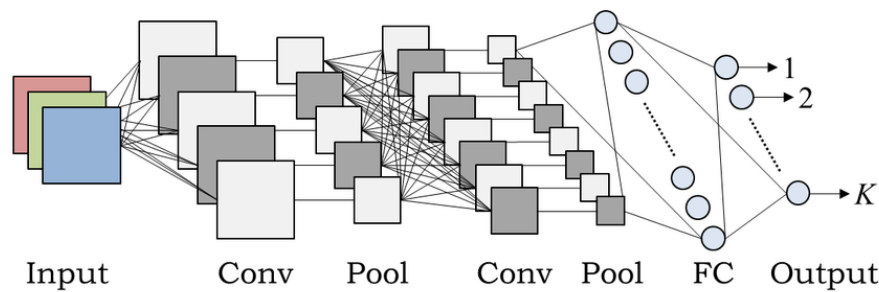

**Figure 1.** *Basic CNN Architecture*

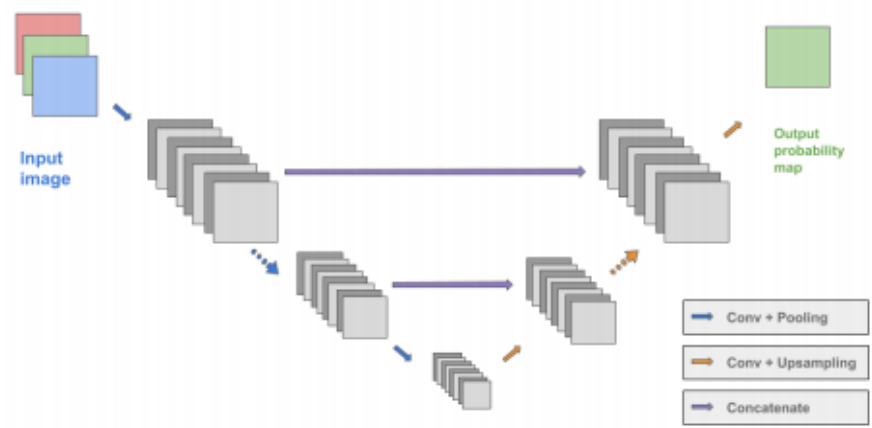

**Figure 2.** *FCA Architecture*

### 3.3. Contrast Limited Adaptive Histogram Equalization (CLAHE)

In histogram equalization, which is frequently used in image enhancement, image quality can be improved by expanding the density dynamic range with the entire image histogram. In histogram equalization, the intensity distribution of the image is normalized to obtain the result image with a uniform intensity distribution, and thus the improvement process is performed.

However, since histogram equalization uses the density distribution of the entire image, this can cause a faded effect on some images when the average intensity is set to medium. And in images with a crowded density distribution in a narrow region, it can cause many noise pixels to occur. Local histogram equalization techniques have been developed to solve these problems [18].

Adaptive histogram equalization is a modified histogram equalization operation and performs optimization on local data. The main idea here is that the image is divided into rectangular regions in the form of a grid and standard histogram equalization is applied to each region. Optimal region sizes and number vary depending on the image.

After the image is divided into sub-regions and histogram equalization is applied to each region, the sub-regions are combined with the bi-linear interpolation method to obtain an improved whole image [19]. However, the noise problem arose in adaptive histogram equalization. To prevent this, it is necessary to limit contrast enhancement in homogeneous regions, and for this purpose, the contrast limited adaptive histogram equalization (CLAHE) method has been developed.

The image is divided into small blocks called tiles. AHE equalizes the histogram for each tile as well. If there is some noise on the floor, the AHE amplifies the noise. Therefore, contrast limitation is used to avoid this problem. If there are any histogram splits above the contrast limit, these pixels are cut and evenly distributed. After equalization, linear interpolation is applied to correct the artificiality of the tile borders. With the CLAHE method, it can improve contrast in medical images, foggy images, underwater images, satellite images and natural images without increasing the effect of noise.

## 4. Discussion and Results

The manual process of histopathological analysis is laborious, time-consuming, and limited by the quality of the specimen and the experience of the pathologist. Exactly for this reason, the aim of the thesis will be to carry out this process in a computer environment, and an important step will be taken to eliminate the different decisions both in terms of time and between pathologists.

The dataset to be used in this study will be studied on the early diagnosis of Paratuberculosis (pTB) disease that has no treatment/vaccine, which is frequently seen in cattle / small cattle in the Burdur region.

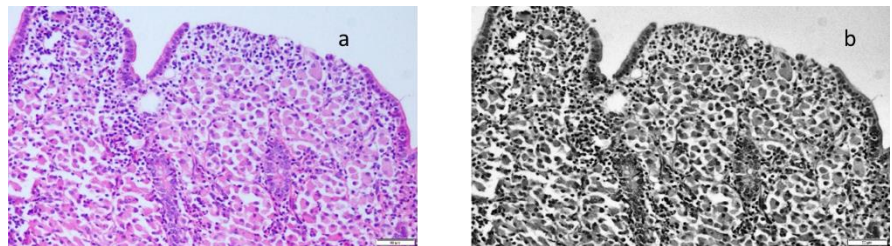

**Figure 3.** *a) Paratuberculosis image b) after data preprocessing with CLAHE*

In this study, the Fastai library was used. Fastai is a deep learning library built on PyTorch, developed to simplify the deep neural networks training process using cutting-edge deep learning approaches in various fields (e.g. computer vision, natural language processing, etc.).

Fastai is a modern deep learning library built on two main designs. It turns out to be a deeply configurable library as well as being accessible and fast productive. This library preserves both the clarity and development speed of the Keras library and the customization ability of the Pytorch library.

Recommended Fastai/VGG16 Model; It is a simple network model and the most important difference from the previous models is the use of convolution layers in 2 or 3 layers. It is converted into a feature vector with 7x7x512=4096 neurons in the full link (FC) layer. The 1000 class softmax performance is calculated at the output of the two fully connected layers. Approximately 140 million parameter calculations are made. As in other models, while the height and width dimensions of the matrices decrease from the input to the output, the depth value (number of channels) increases. At each convolution layer output of the model, filters with different weights are calculated, and as the number of layers increases, the features formed in the filters represent the 'depths' of the image.

The major contribution of the proposed Fastai/VGG-16 network shows that the depth of the network is a critical component for good performance. In this architecture, from the beginning to the end, only the 2x2 size filter is used in the convolution process, while keeping the network simple and protected in depth.

Although the network's biggest drawback was that it needed a lot of memory and parameters (approximately 140 million) when it was first proposed, it has been determined by the study that most of these parameters are in the first fully connected layer and removing this layer does not decrease the performance of the network much.

In this study, the effect of CLAHE method, which is one of the data preprocessing techniques, on histopathology images using deep learning was investigated. Metrics in the classification made on the data set without and using the CLAHE method; Classification Accuracy, Sensitivity, Specificity, Precision and F1 Scoring were performed. Fastai/VGG16 values without using the CLAHE method; shown in the table1below.

**Table 1.** *Performance Metrics (without CLAHE Method)*

| Confusion Matrix Results VGG16 | Classification Accuracy | Sensitivity | Specificity | Precision | F1 SCORE |
|---|---|---|---|---|---|
| TP | 40 | 0,933333333 | 0,952381 | 0,9166667 | 0,909091 | 0,9302326 |
| TN | 44 | | | | | |
| FP | 4 | | | | | |
| FN | 2 | | | | | |

Values using Fastai/VGG16 with CLAHE method; shown in table 2.

**Table 2.** *Performance Metrics of Proposed Model Fastai/VGG16 (with CLAHE Method)*

| Confusion Matrix Results VGG16 | Classification Accuracy | Sensitivity | Specificity | Precision | F1 SCORE |
|---|---|---|---|---|---|
| TP | 90 | 0,983606557 | 0,97826087 | 0,989011 | 0,989011 | 0,9836066 |
| TN | 90 | | | | | |
| FP | 1 | | | | | |
| FN | 2 | | | | | |

## 5. Conclusions and Future Work

As a result of the study, in this context; the use of deep learning systems for pathological diagnosis is a very new technique. However, it is rapidly becoming widespread due to the high rate of accurate and rapid diagnosis. This study showed that deep learning can be used to diagnose paratuberculosis.

In this study, a diagnosis and classification method was developed for paratuberculosis disease, which affects both animals and humans, using a new and original data set.

The data set of this study is expected to set an example for those working on disease classification and diagnosis. The data set to be shared with the employees on this subject may allow the development of different models.

**Declaration of interest**

It was presented at the ICAIAME 2021 conference and published as a summary.